# Advancements in Data Processing and Calibration for the Hyperspectral Imaging Satellite (HySIS)

Ankur Garg, Abhishek Patil, Meenakshi Sarkar, S. Manthira Moorthi, Debajyoti Dhar [*]

October 23, 2024

**Abstract**

Hyperspectral imaging stands as a powerful tool in Earth exploration, owing to its exceptional ability to discern intricate spectral features. India has harnessed this capability through the launch of a dedicated hyperspectral earth observation satellite. This satellite is designed to scrutinize the Earth's surface across a broad electromagnetic spectrum range spanning from 400 nm to 2500 nm, capturing an extensive array of spectral information. One of the key strengths of this satellite lies in its impressive spatial sampling resolution of 30 meters. This high resolution further amplifies its imaging prowess, enabling it to capture detailed and accurate data about the Earth's surface features. This paper offers a comprehensive account of the meticulous calibration procedures that were meticulously executed during the pre-launch phase. Calibration is a critical process in ensuring the accuracy and reliability of the data captured by the satellite. This involved a series of precise adjustments and measurements to fine-tune the satellite's imaging instruments, ensuring that they were operating at their optimal levels. The paper also highlights the achieved performance metrics, showcasing the satellite's capability to deliver high-quality, detailed imagery of the Earth's surface. These metrics serve as a testament to the effectiveness of the calibration procedures carried out prior to the satellite's launch. Furthermore, the paper outlines the rigorous in-orbit radiometric and geometric calibration and validation endeavors that were conducted during the commissioning phase of the mission. This phase involved further fine-tuning and validation of the satellite's instruments in the actual space environment. It ensured that the satellite was operating accurately and reliably in orbit, ready to fulfill its mission objectives. In addition, the paper sheds light on the observations made and anomalies encountered during the in-orbit phase. These observations provide valuable insights into the challenges faced in space and serve as a foundation for the development of specialized algorithms. These algorithms were specifically devised to address and correct any anomalies encountered, further enhancing the satellite's operational efficiency and data quality. The operational data processing pipeline is also delineated in detail within the paper. This encompasses the pivotal stages involved in transforming raw satellite data into usable, high-fidelity Earth observations. This includes data acquisition, pre-processing, calibration, and final image generation. Finally, the paper furnishes a thorough assessment of the radiometric and geometric quality achieved up until the present date. This assessment serves as a testament to the satellite's operational efficiency and reliability in delivering high-fidelity Earth observations. It provides valuable insights into the long-term performance and sustainability of the satellite's mission, reinforcing its significance in advancing Earth exploration through hyperspectral imaging technology.

# 1 Introduction

The PSLV-C43 mission marked a significant milestone in space exploration, launching the Indian Hyperspectral Spectrometer as its primary payload on November 29, 2018, at 09:57 hrs (IST). This remarkable event took place at the First Launch Pad (FLP) at the Satish Dhawan Space Centre in Sriharikota, under the auspices of the esteemed Indian Space Research Organisation (ISRO). The satellite's principal objective was to conduct detailed studies of the Earth's surface within the visible, near infrared, and shortwave infrared segments of the electromagnetic spectrum, harnessing the power of hyperspectral imaging.

[†]Corresponding author : agarg@sac.isro.gov.in

[*]All authors are with Signal and Image Processing Group, Space Applications Center, Indian Space Research Organisation, Ahmedabad, India-380015

Table 1: Mission Specifications

| **Mission Specifications** | | |
|---|---|---|
| | VNIR | SWIR |
| Imaging Principle | Pushbroom-Grating | |
| Spectral Range | 400-900 nm | 850-2500 nm |
| Spectral Sampling | ∼10 nm | ∼6.5 nm |
| Spectral Resolution / Bandwidth | ∼8-10 nm | ∼9-10 nm |
| Spectral channels | 60 | 256 |
| Spatial Samples | 1000 samples | |
| Radiometric Resolution | 12 bits | |
| Ground Sampling Distance | 30 m in nadir | |
| Orbit Specification | 630 km Polar sun-synchronous near circular orbit, Step & Stare Ratio of 5:1 | |

Table 1 provides a comprehensive overview of the key mission and payload parameters, encapsulating critical details about the satellite's specifications and capabilities. Additionally, Figure 1 visually portrays the integrated spacecraft, showcasing the specialized VNIR (Visible and Near Infrared) and SWIR (Shortwave Infrared) instruments that constitute its imaging system.

An indispensable facet of satellite data processing involves the rigorous modeling of both the imaging system, which encompasses the instrument and platform characteristics, and the rectification of any systematic or unsystematic errors that may impact radiometric and geometric performance. This process is pivotal in ensuring the accuracy and reliability of the data generated. Moreover, it strives to cater to the needs of application users by presenting the data in readily analyzable formats.

Precise characterization and calibration of any scientific instrument are paramount for accurately converting acquired data from digital to physical units. This monumental endeavor commences with instrument calibration in a controlled lab environment, where meticulous measurements and adjustments are made. The process then advances to the comparison and evaluation of radiometric and geometric performance using in-orbit data, ensuring that the instrument operates optimally in the space environment. This culminates in the development of correction methodologies for any identified anomalies, thereby refining the quality of the captured data.

Processing hyperspectral data for radiometric and geometric fidelity poses unique challenges compared to handling panchromatic and multispectral data. This is attributed to the multitude of spectral channels, finer spectral resolution, the presence of keystone and smile errors, and the utilization of multiple instruments to cover the requisite spectral range. This paper provides a comprehensive overview of the methodologies employed for both on-ground lab characterization and in-flight data calibration and validation for the Indian spectrometer. Furthermore, it expounds upon the sophisticated algorithms implemented to enhance radiometric and geometric performance, addressing the specific challenges posed by hyperspectral imaging technology. Through these endeavors, the satellite ensures that it delivers data of the highest quality and accuracy, thereby maximizing its potential for valuable Earth exploration insights.

## 2 Radiometric Lab Characterization

Remote sensing studies rely on the extraction of geophysical parameters from the digital readings produced by the instruments on board a satellite. These digital readings are then transformed into meaningful geophysical units such as radiance or reflectance, a conversion process grounded in the relationships established during the instrument's calibration. In the lead-up to launch, each remote sensing instrument undergoes a rigorous regimen of characterization and calibration protocols conducted within a tightly controlled laboratory environment [1],[2],[3]. This all-encompassing process encompasses an in-depth examination of various crucial parameters including the instrument's flat-field response, signal-to-noise ratio performance, saturation radiance levels, central wavelengths, bandwidths, relative spectral response (RSR), Square Wave Response (SWR), as well as assessment for smile and keystone errors. These measurements hold paramount importance, serving as the gold standard for evaluating the instrument's anticipated performance once it is deployed in orbit. They establish a baseline against which the instrument's actual performance can be compared, allowing for any deviations or discrepancies to be identified and addressed. Upon deployment in orbit, a meticulous and precise comparison of the instrument's real-world performance with the predefined benchmark values is conducted. Any observed deviations are scrutinized using sophisticated modeling techniques. These techniques are employed to pinpoint the source of the deviation and to rectify it, thereby ensuring that the instrument's data output remains consistently accurate and reliable for subsequent remote sensing endeavors. In essence, the calibration process represents a critical phase in the lifecycle of a remote sensing instrument. It is the linchpin that bridges the raw digital counts to scientifically meaningful geophysical parameters. Through the careful execution of characterization and calibration procedures, the instrument's performance is rigorously assessed, verified, and fine-tuned, ultimately ensuring the integrity and precision of the data it produces for a wide range of remote sensing applications. This meticulous approach is essential for extracting reliable and scientifically valuable information from the vast array of data collected during satellite missions.

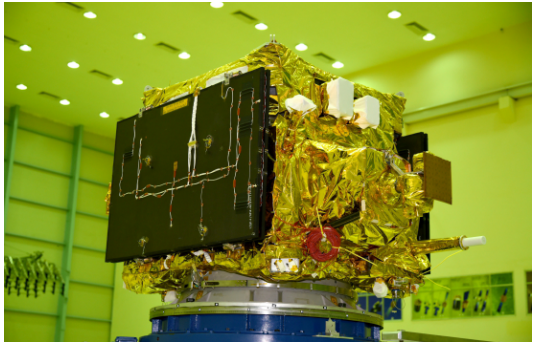

Figure 1: Assembled Spacecraft (Source : ISRO Website)

## 2.1 Spectral Calibration

Accurate knowledge of an instrument's spectral response is of paramount importance for conducting rigorous scientific studies. Prior to launch, the instrument undergoes a meticulous process of spectral calibration to precisely measure crucial parameters. These include the central wavelength, bandwidth, and Relative Spectral Response (RSR) for all channels across the field of view (FOV).

The RSR, a pivotal function, serves a dual role. It not only plays a crucial role in the lab transfer characterization of the instrument but also holds great significance in the atmospheric correction of acquired data. Essentially, RSR represents the pixel's response when subjected to monochromatic light of varying wavelengths spanning the entire electromagnetic spectrum. During the spectral calibration process, the shape of the RSR is established as an integral part of characterizing the instrument's behavior. The central wavelength of a channel is determined as the peak location of the RSR, while the bandwidth, also known as the Full Width Half Maximum (FWHM), is the difference between the locations where the response reaches 50

For both the VNIR and SWIR instruments, as part of the calibration activity, they were subjected to monochromatic light with a bandwidth of 2 nm and a 1 nm sampling rate, covering the spectrum from 350 nm to 2600 nm. This rigorous process ensures that the instruments are precisely calibrated to capture data across this wide spectral range.

The concept of a spectral smile or frown curve is crucial in understanding the instrument's behavior. It denotes the shift in central wavelength across the FOV for a spectral channel [4]. A smile artifact can significantly impact the accuracy of geophysical parameter retrieval, reduce classification precision, and introduce ambiguity in determining absorption features. The smile phenomenon is typically induced by two primary factors: optical aberrations across the FOV or in-plane rotation of the 2D detector [3]. In cases where smile is caused by in-plane detector rotation, the variation is linear, whereas for smile induced by optical aberrations, the variation is non-linear.

Figure 2 visually represents the alterations in central wavelength variation across the FOV for both the VNIR and SWIR spectrometers, along with the FWHM response. On average, the FWHM was measured to be approximately 9.24 nm for VNIR and 5.87 nm for the SWIR instrument (as depicted in Figure 2), both of which fall within the specified parameters. The smile observed in the VNIR spectrometer is confined within a single pixel, exhibits non-linear behavior, and is primarily attributed to optical aberration. Conversely, the smile present in the SWIR spectrometer spans approximately 2 pixels, displays linear behavior, and is predominantly influenced by in-plane detector rotation. These detailed characterizations and calibrations are essential for ensuring the accuracy and reliability of the instrument's data output in a wide range of remote sensing applications. They serve as the foundation for precise scientific analysis and interpretation of Earth observation data.

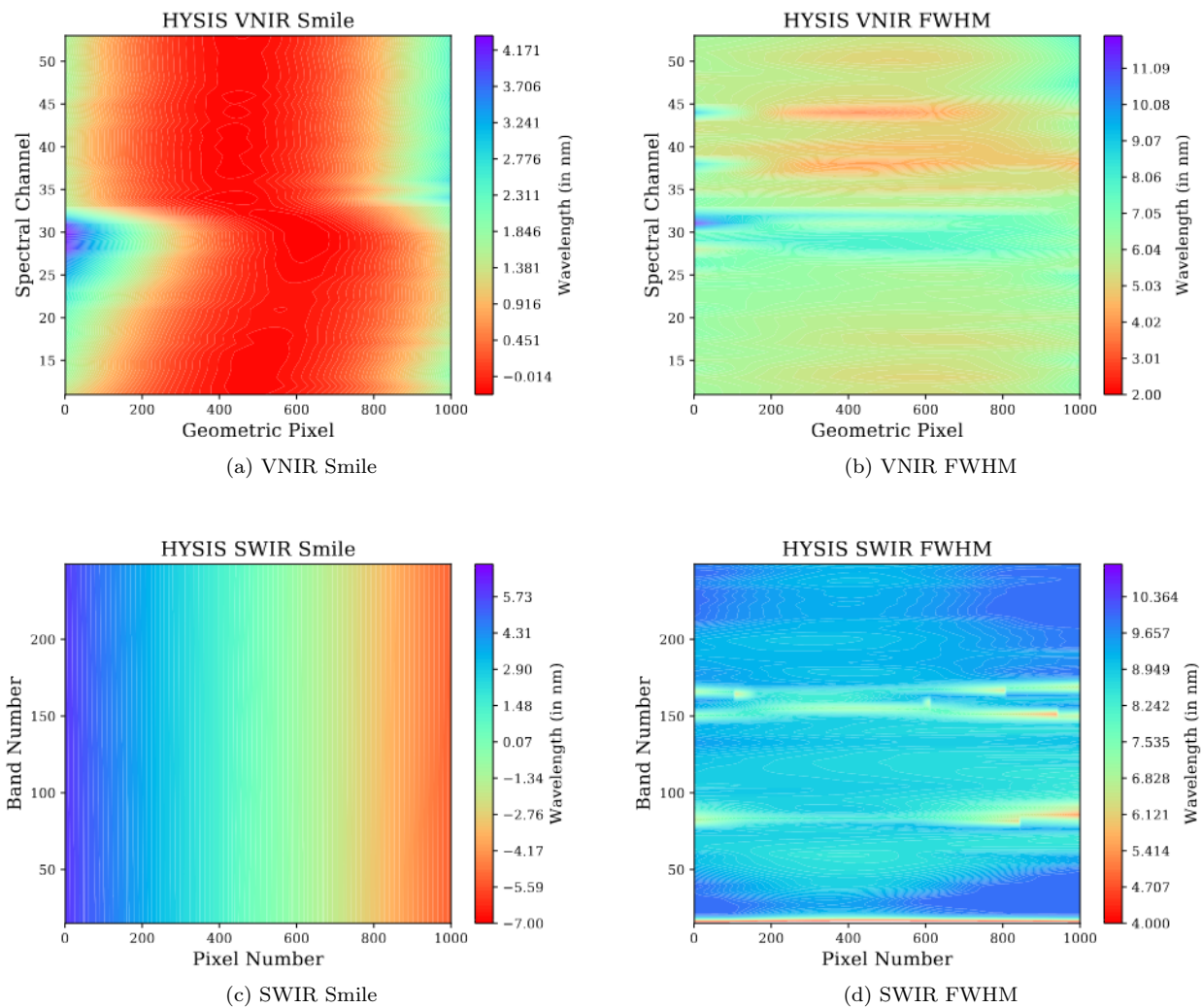

(a) VNIR Smile  (b) VNIR FWHM
(c) SWIR Smile  (d) SWIR FWHM

Figure 2: Spectral Calibration Results

## 2.2 Light Transfer Characterization / Flat Field Response

In the realm of remote sensing, the measurement of physical quantities differs from traditional ground-based methods. Rather than directly assessing these quantities, they are derived from observations made by satellites. The sensor aboard the satellite records digital counts, and a critical step in the calibration process is establishing a relationship between these counts and the received radiance [1].

One pivotal facet of this calibration is flat-field calibration. This intricate procedure involves exposing each pixel in every spectral channel of the instrument to diffused light emanating from an integrating sphere, which functions as a uniform lambertian source. The instruments are illuminated with varying intensities of this uniform light, and their responses are meticulously recorded. The resulting dataset is then harnessed to establish the conversion from digital counts to meaningful physical units. Additionally, it serves as the foundation for generating flat-field correction coefficients, which play a vital role in normalizing pixel responses across the entire field of view (FOV).

As part of the flat-field calibration exercise, the Signal-to-Noise Ratio (SNR) and saturation radiance for different channels are thoroughly evaluated. Figures 3 and 4 vividly depict the instrument's response both before and after flat-field correction for the central channel. It is evident that the non-uniformities present in the initial response are significantly reduced, often to less than 1

The Signal-to-Noise Ratio (SNR) at 100

In summary, flat-field calibration represents a crucial step in the process of converting digital counts to physically meaningful measurements in remote sensing. It plays a fundamental role in ensuring the accuracy and reliability of the data collected by the instrument. Through meticulous calibration procedures, the instrument is fine-tuned to deliver precise and consistent measurements, enabling scientists and researchers to extract valuable insights from satellite observations.

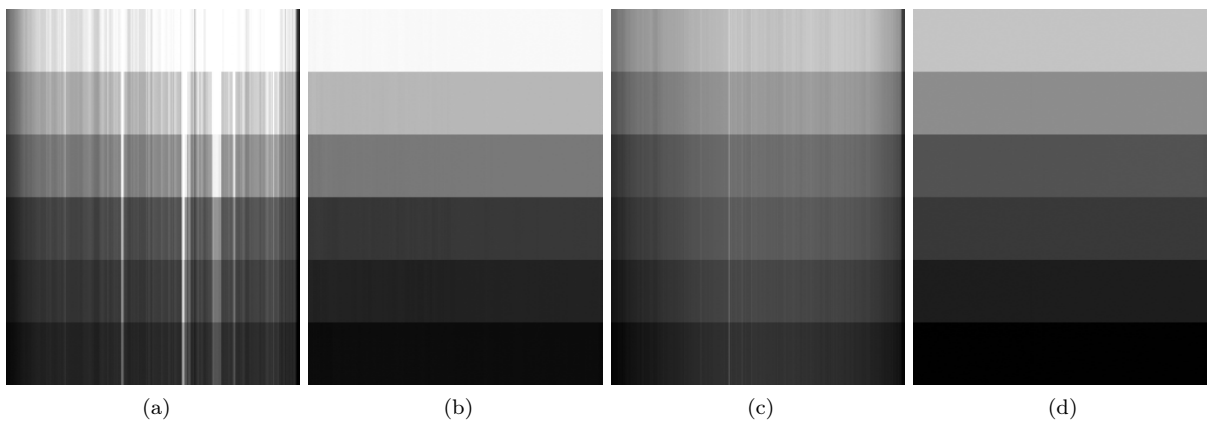

(a)      (b)      (c)      (d)

Figure 3: Flat Field Responses (a) VNIR (Band30 Before) (b) VNIR (Band30 After) (c) SWIR (Band128 Before) (d) SWIR (Band128 After)

## 2.3 Keystone Characterization

The keystone artifact in a hyperspectral spectrometer refers to a spatial mis-registration caused by variations in magnification that depend on wavelength across different spectral channels [5]. This artifact has a significant impact on the acquired data, as it leads to a shift in the spatial location of features and ultimately results in a distorted or corrupted spectrum. The keystone artifact can originate from two primary sources: chromatic aberration and in-plane detector rotation. Chromatic aberration occurs due to the wavelength-dependent focusing of light, leading to variations in magnification across the spectral range. In-plane detector rotation, on the other hand, arises from misalignment or rotation of the detector relative to the optical axis, causing a spatial shift in the recorded image. To quantitatively assess the extent of the keystone present in the instrument, bar targets are imaged for all spectral channels. These targets serve as a reference to evaluate the spatial alignment across the spectral range. In Figure 6, the estimated keystone for the VNIR spectrometer's central field is displayed. It reveals that a keystone artifact with a magnitude of approximately ±1.5 pixels is present in the VNIR instrument. This means that there is a spatial mis-registration of approximately 1.5 pixels between different spectral channels, which can have a noticeable impact on the accuracy of the spectral data. Addressing and correcting for the keystone artifact is crucial for ensuring the integrity and accuracy of hyperspectral data. By understanding the sources and magnitude of this artifact, corrective measures can be implemented to improve the spatial registration of features across different spectral channels, ultimately enhancing the quality and reliability of the acquired data for scientific analysis and interpretation.

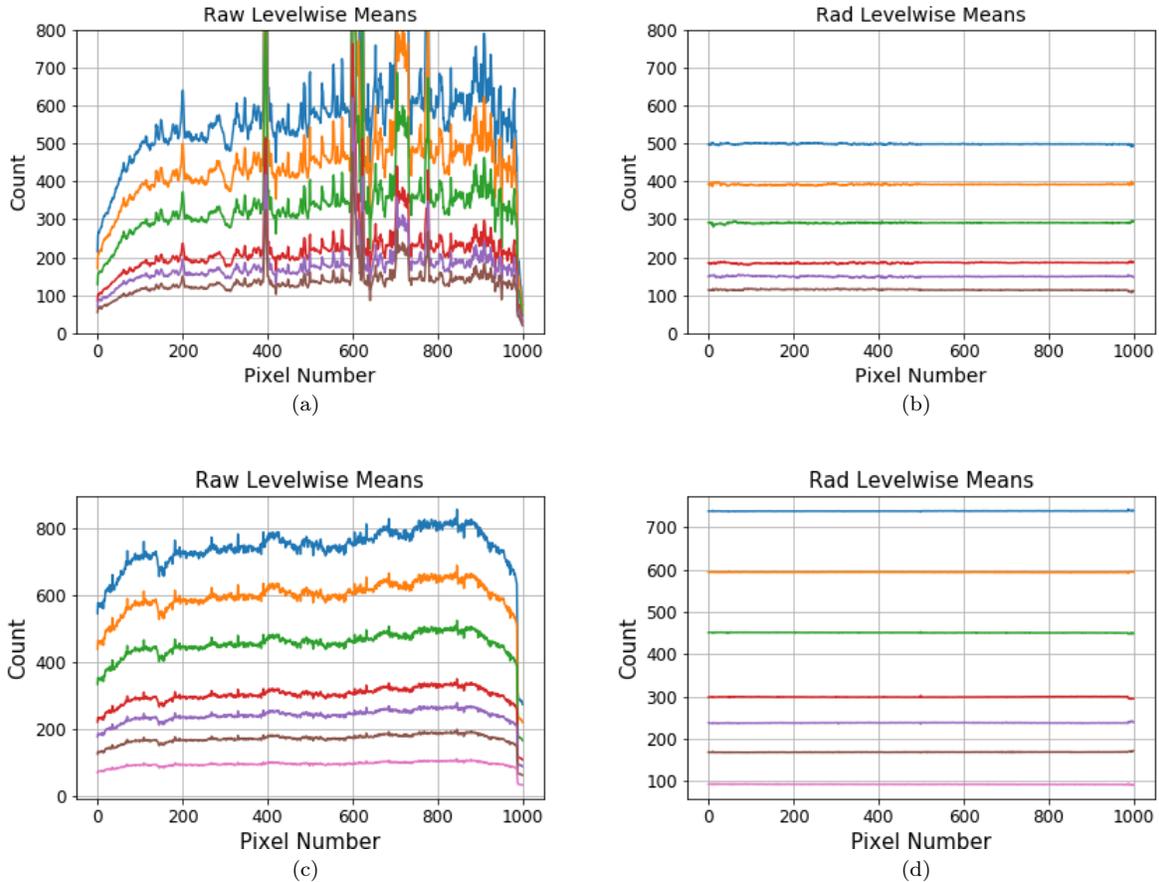

Figure 4: PRNU Correction Performance (a) VNIR (Band 30 Before) (b) VNIR (Band 30 After) (c) SWIR (Band 128 Before) (d) SWIR (Band 128 After)

## 3 In-flight Radiometric Characterization

During the commissioning phase of a mission, it is crucial to conduct a rigorous reevaluation and validation of the sensor's performance against ground-based measurements. This phase serves as a pivotal checkpoint to ensure that the sensor is operating as expected in the space environment. Any deviations or discrepancies observed between the sensor's actual performance and its expected ground performance are meticulously scrutinized. This careful examination helps identify potential issues or anomalies that may have arisen during the transition from the controlled laboratory environment to the real-world conditions of space. Based on these findings, necessary adjustments are made to the data processing algorithms to correct for any discrepancies and fine-tune the sensor's performance. Furthermore, comprehensive datasets covering the full dynamic range of the sensor are acquired during this phase. These datasets encompass a wide range of intensities and conditions to thoroughly calibrate and validate various system parameters. This process involves exposing the sensor to diverse scenarios, including different platform biases and rates, to ensure that it accurately captures and processes data under varying conditions. As part of the commissioning phase, a detailed reassessment of all significant system parameters is conducted. This includes but is not limited to flat-field correction coefficients, saturation radiance levels, Signal-to-Noise Ratio (SNR), spectral shifts, keystone artifacts, and geometric biases. Each parameter is carefully evaluated to ensure that it aligns with the predefined specifications and meets the required performance standards. The recalibrations and validations carried out during the commissioning phase represent critical steps in the mission's overall success. They are instrumental in ensuring the accuracy, reliability, and quality of the data collected during the operational phase of the mission. By fine-tuning and confirming the sensor's performance in space, these efforts lay the foundation for a successful and scientifically valuable mission, enabling researchers and scientists to derive meaningful insights from the acquired data.

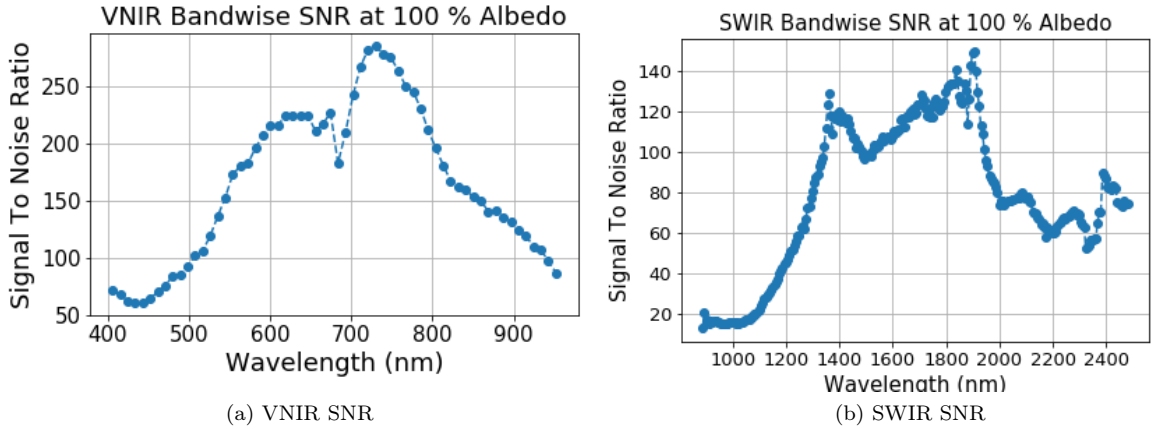

(a) VNIR SNR
(b) SWIR SNR

Figure 5: Signal To Noise Ratio for VNIR & SWIR

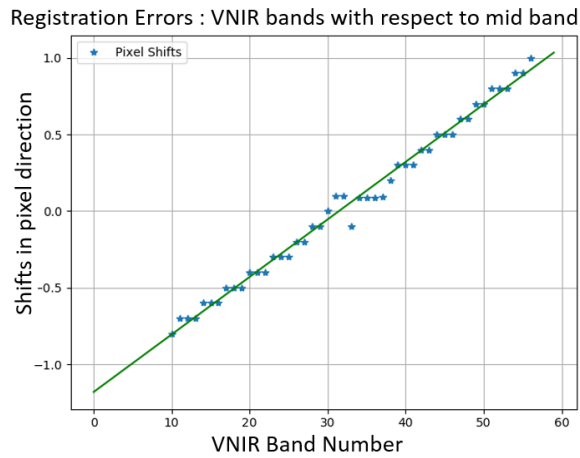

Figure 6: Keystone Estimation for VNIR Sensor

## 3.1 Dark Bias Estimation

- VNIR : The VNIR instrument's detector encompasses 30 channels that do not receive any illumination. These unilluminated pixels provide a reference for what is known as dark biases. These dark biases represent inherent electronic noise or baseline signal that may be present in the data acquired from the illuminated channels. By characterizing and quantifying these dark biases, it becomes possible to make appropriate adjustments in the signals received from the illuminated channels. This process is crucial in ensuring that the data accurately reflects the true radiometric values of the observed targets. It effectively mitigates any potential biases introduced by the instrument's electronics.

- SWIR : The SWIR instrument exhibits a notable sensitivity to variations in operating temperature. This sensitivity necessitates correction for two specific factors: dark current and background noise. Unlike the VNIR instrument, the SWIR instrument lacks unilluminated pixels that can serve as a reference for dark biases. To address this, a comprehensive data collection effort was undertaken during the initial in-orbit testing phase. Dark data was collected over a period of 10 consecutive days for both instruments. The trends observed in this data were systematically analyzed and monitored in relation to variations in operating temperature. It was discerned that these trends exhibited a remarkable level of consistency over time. The dark bias information derived from this analysis was subsequently employed to effectively adjust the saturation radiance levels in the SWIR instrument. This correction process is essential in ensuring that the data accurately represents the spectral characteristics of the observed targets. By eliminating unwanted noise and biasing effects, the corrected data provides a more faithful representation of the true radiometric values. In summary, both the VNIR and SWIR instruments employ sophisticated correction methodologies tailored to their specific characteristics and challenges. These corrections are vital in guaranteeing that the data generated by these instruments is of the highest quality and fidelity, enabling accurate and reliable scientific analysis across a broad range of applications.

## 3.2 Flat Field Correction Coefficients Updation

In the harsh conditions of orbit, the response of pixels in a satellite sensor can undergo changes over time. Factors like shifts in operating temperature and natural detector aging contribute to these alterations. To counteract these deviations and ensure uniformity in the detector's performance, it becomes imperative to recalibrate the flat-field correction coefficients using in-orbit data. This recalibration process involves capturing images of uniform targets in diverse environments, including desert, water, snow, and deep space. These targets cover a wide range of intensities and conditions, encompassing the full dynamic range of the sensor. From these in-orbit observations, new flat field coefficients are derived to account for any changes in the sensor's response. As illustrated in Figure ??, the datasets utilized for updating coefficients for both instruments are displayed. Figure 7 further demonstrates the impact of this recalibration process, showcasing images of desert and water using the older and newly updated correction coefficients. The comparison reveals a significant reduction in non-uniformity, decreasing from approximately 8% to about 2%, with the implementation of the new coefficients. This remarkable improvement in uniformity underscores the effectiveness of the recalibration process. This recalibration endeavor is instrumental in upholding the accuracy and reliability of the sensor's performance over the course of the mission. By continuously monitoring and adjusting for any changes that may occur during the mission's operational phase, this process ensures that the sensor consistently delivers high-quality and accurate data. This, in turn, enables researchers and scientists to derive meaningful and reliable insights from the acquired data throughout the duration of the mission.

## 3.3 In-flight Spectral Calibration and Validation

The smile effect, resulting from detector in-plane rotation and chromatic aberrations, can experience variations due to shifts in temperature and mechanical stresses during launch. Therefore, it is crucial to validate and recalibrate the smile effect observed using in-orbit data. This process ensures that the instrument's performance remains accurate and reliable in the space environment. To accomplish this, data strips containing the same target across the field of view (FOV) are carefully selected. A one-dimensional sub-pixel phase correlation method [6] is applied to the spectral signatures in relation to the central pixel's signature. This involves using a small one-dimensional window, typically spanning 5 or 10 pixels, across the spectral response of each channel and at each spatial location. Any observed shift in the spectral direction is indicative of smile in the instrument. Additionally, there can be an absolute shift in the measured central wavelengths of channels compared to the actual signature of the target. After correcting for smile effects, the absolute shift in observed absorption features is determined by examining dips caused by $CO_2$, $H_2O$, and $O_2$ absorption lines. For instance, oxygen exhibits absorption dips at approximately 760 nm, 1461 nm, and a weaker dip at around 1265 nm. Water vapor induces absorption dips at approximately 823.04 nm, 936 nm, and 1133 nm. Similarly, $CO_2$ generates a prominent atmospheric dip at approximately 1570 nm, 2010 nm, 2060 nm, and a weaker dip at around 1610 nm. Resolving relative smile necessitates data resampling in the spectral direction, which can vary across the FOV. Conversely, absolute shift does not require such data resampling. For the VNIR instrument, the smile was found to be consistent with the determination made during lab characterization, approximately 4.17 nm across the FOV. Similarly, the smile error for the SWIR instrument matched the lab characterization, at around 12.0 nm across the FOV. The absolute shift in central wavelength was determined to be approximately 5.5 nm for the VNIR instrument and about 9 nm for the SWIR instrument, as depicted in Figure 8. These values align with the lab calibration, providing assurance that the instrument's smile effect has been accurately characterized and corrected for in the in-orbit phase. This meticulous validation and recalibration process play a crucial role in ensuring the accuracy and reliability of the instrument's spectral measurements during the mission's operational phase.

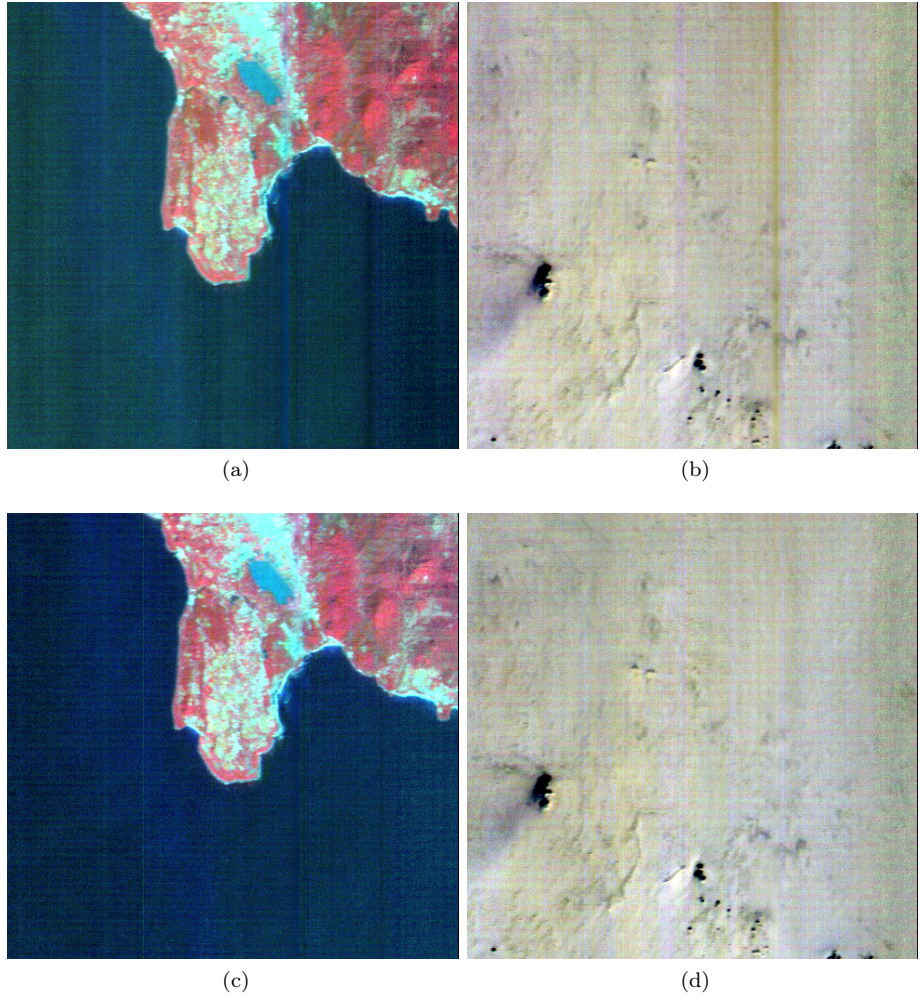

Figure 7: VNIR Samples Showing Improvement in PRNU, Before and After LUT Update

## 3.4 In-flight Vicarious Calibration

Vicarious calibration represents a post-launch calibration approach that involves calibrating sensors by utilizing natural or artificial targets imaged in close proximity to the sensor's observations. This method relies on ground-measured reflectance values, which are then converted to top-of-atmosphere radiance using a radiative transfer model [2]. The radiance values measured by the sensor are subsequently compared to the simulated radiance, facilitating the generation of calibration coefficients. These coefficients play a crucial role in refining the observed spectral signatures of various targets. In Figure 9, a visual representation is provided, showcasing the instrument-measured signatures before refinement, alongside the reference and calibrated signatures. These targets encompass a range of environments, including vegetation, water, sand, grass, and even the moon, as presented in Figure ??. These strips serve as essential data sources for fine-tuning the calibration coefficients. Upon close examination, it is evident that the calibrated signatures closely align with the reference signatures. Before the refinement process, the percentage deviation was found to be as high as approximately 200%. However, through the calibration and refinement process, this deviation was notably reduced to an average of around 10%. It is worth noting that despite the significant improvement, there are a few channels that still exhibit noticeable deviations. To account for this, these channels are marked as "bad" in the metadata, signifying a lower confidence level, and are consequently not recommended for use in applications. Furthermore, plans are underway to incorporate additional ground targets in the near future. These additions will further enhance and fine-tune the calibration coefficients, ultimately leading to even more accurate and reliable spectral measurements. This continuous refinement process ensures that the sensor's observations remain precise and trustworthy, providing valuable data for a wide range of applications in remote sensing and Earth observation.

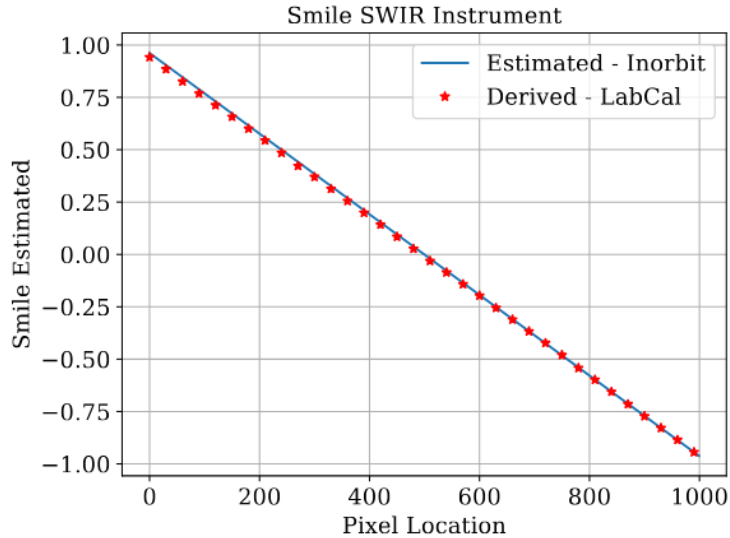

(a)

Figure 8: Validation of In-orbit Estimated Smile with Lab Data For SWIR Instrument

## 3.5 In-flight Keystone Characterization

As mentioned earlier, the keystone error leads to spatial mis-registration among different spectral bands. In orbit, this error was evaluated by calculating sub-pixel shifts across the field of view (FOV) for all spectral bands relative to channel 30, which serves as the central spectral channel in the VNIR instrument. Figure ?? visually represents the pixel-to-pixel error across bands, determined using this method. The observed value was found to be consistent with the value determined through laboratory calibration. However, it's important to note that for the SWIR instrument, the validation of the keystone error was not feasible due to the presence of stray light, which interfered with the estimation process. This highlights the need for further efforts or alternative methods to accurately assess the keystone error for the SWIR instrument in orbit. The presence of stray light in the SWIR instrument introduces a unique challenge that must be carefully addressed to ensure accurate and reliable measurements. Strategies such as advanced modeling techniques or specialized calibration procedures may be required to effectively characterize and correct for the keystone error in the SWIR instrument during the mission's operational phase. Continued research and development in this area will be crucial for maximizing the scientific value and precision of the data collected by the satellite, particularly in the SWIR spectral range. It underscores the dynamic nature of space-based instrumentation and the importance of adaptability and innovation in maintaining the integrity of scientific measurements in challenging operational environments.

# 4 Radiometric Data Processing Chain

The operational data processing chain serves as a critical component in the generation of data products, ensuring they exhibit high levels of radiometric and geometric accuracy while adhering to the prescribed format. This is achieved through the application of a diverse array of algorithms tailored to address specific issues encountered during data analysis. These correction methodologies are carefully implemented to enhance the overall quality and reliability of the resulting data products. This systematic approach enables the satellite mission to yield precise and trustworthy information for a wide range of applications, from environmental monitoring to scientific research and beyond.

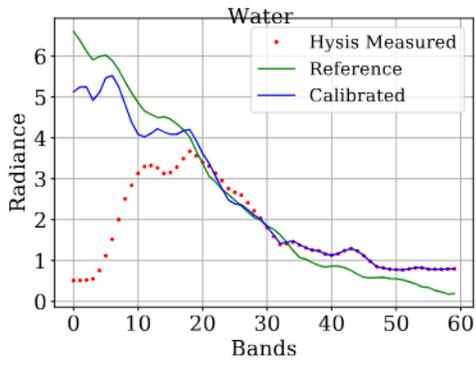
(a)

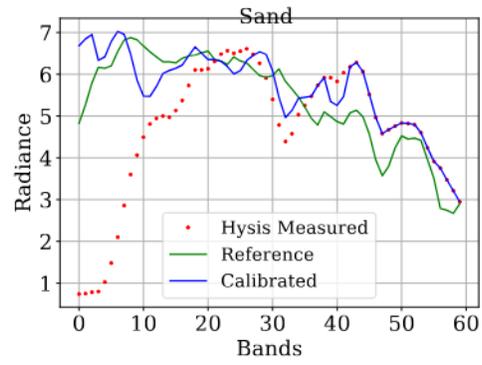
(b)

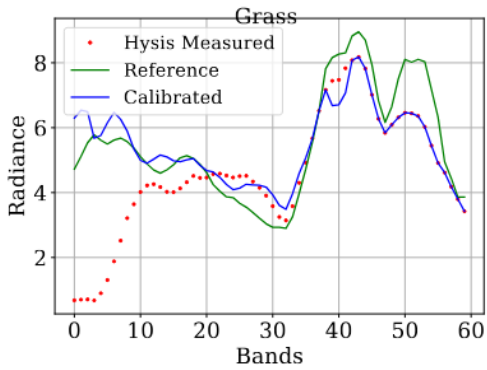
(c)

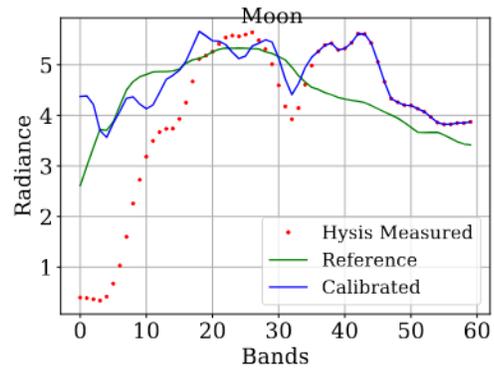
(d)

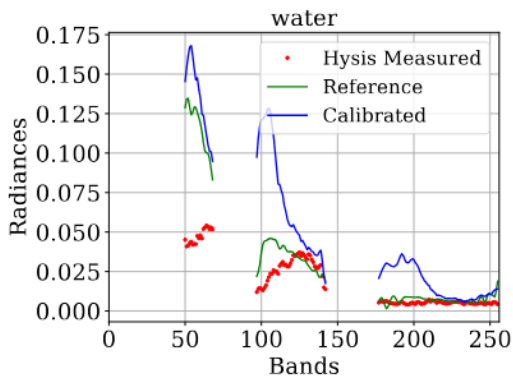
(e)

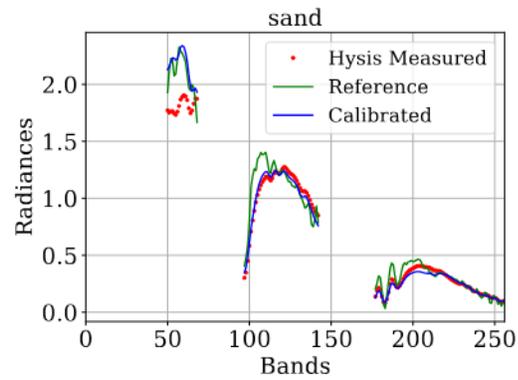
(f)

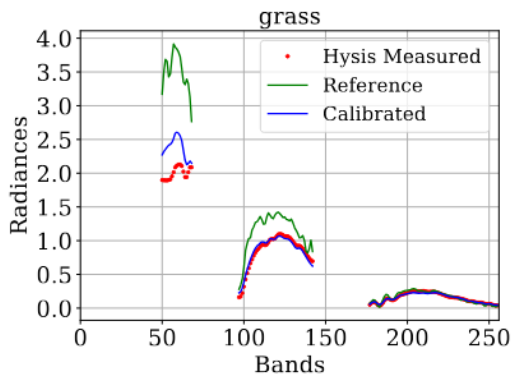
(g)

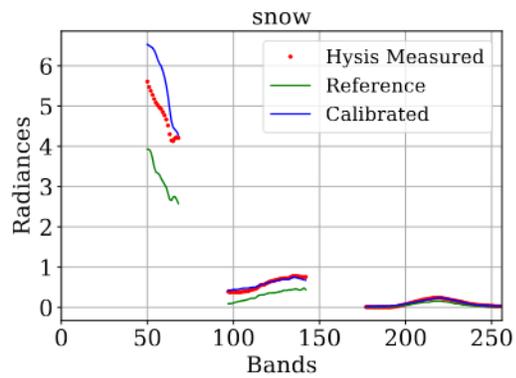
(h)

Figure 9: Spectral Profiles for Water, Sand, Grass, Moon - Before and After Vicarious Calibration for VNIR(a,b,c,d) and SWIR(e,f,g,h)

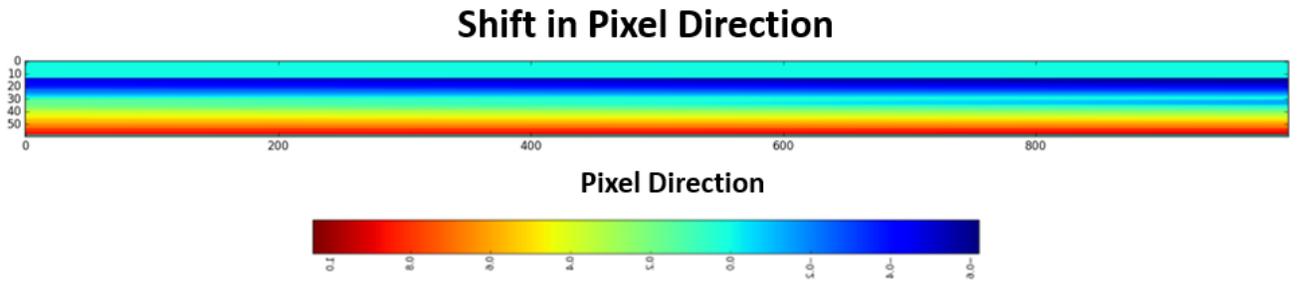

Figure 10: VNIR Estimated Keystone Error from In-orbit Data

## 4.1 Bunch Pixel Loss Correction

The VNIR instrument encountered an issue wherein approximately 10 channels exhibited consecutive pixels, with some extending up to 15, demonstrating significantly elevated responses. These anomalies displayed highly non-linear behavior at three distinct locations across the swath. This phenomenon introduced noticeable artifacts in the captured images, as exemplified in Figure 11 (a) and (c). To address these anomalous pixels, a novel approach was developed and seamlessly integrated into the data processing engine. This innovative method aimed to systematically identify and rectify the problematic pixels. For each affected pixel, a mapping process was initiated to pinpoint the most correlated channel where the same pixel exhibited normal behavior. Once the reference channel was successfully determined, the spectral relationship between the channel requiring correction and the reference band was established for the neighboring pixels surrounding the affected one. Subsequently, the affected pixel was replaced with a corrected value derived from this relationship. Figure 11 provides a visual representation of the original and corrected images, showcasing the successful implementation of this algorithm in desert regions. It is worth noting that this algorithm underwent extensive testing across various target types, demonstrating the anticipated performance and effectiveness in addressing and mitigating the issue of anomalous pixels. As a result, this approach ensures the production of imagery that is accurate, reliable, and free from the artifacts that were initially present. This innovative correction methodology represents a significant advancement in ensuring the quality and integrity of imagery produced by the VNIR instrument, ultimately enhancing the scientific value and utility of the data for a wide range of applications in Earth observation and remote sensing.

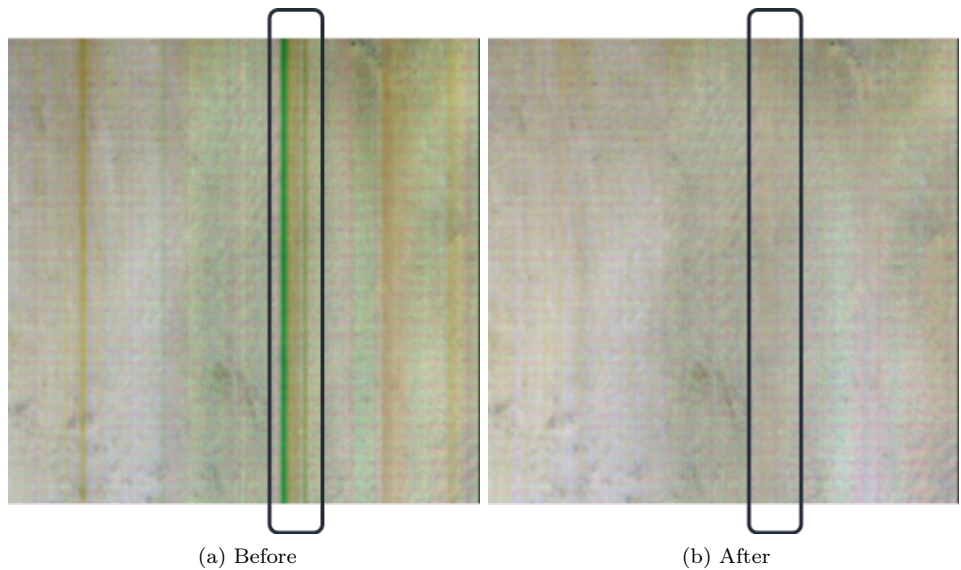

(a) Before  (b) After

Figure 11: VNIR Instrument : Bunch Loss Correction Improvement In Data Usability

## 4.2 Scan Interference Correction

In the SWIR data, distinct horizontal patterns with varying spatial frequencies were identified, exerting a significant influence on the overall image quality. These patterns encompassed both high frequency components, contributing to severe noise, and low frequency components, which manifested as noticeable across-track banding. To address the high frequency periodic patterns, a sophisticated approach involving Butterworth notch filters was employed. These filters were strategically applied to specifically target and attenuate the affected frequencies, effectively suppressing the high frequency noise. For the low frequency banding, spectral channels 160 to 170, situated within the atmospheric absorption window, were selected. The responses from these channels were averaged and subsequently employed for correction across all channels. This approach helped to counteract the low frequency banding artifacts. To further refine the correction process, any residual interference was meticulously dealt with. This was achieved by computing the across-track mean profile for the entire image. The overall mean of this profile was then subtracted from the learned profile, effectively removing any lingering interference. Finally, any remaining interference pattern was subtracted from the image, resulting in a cleaned version with significantly improved quality. Figure 12 visually demonstrates the transformation of channel 100, showcasing its raw state and the substantial improvements achieved after undergoing non-uniformity and scan pattern correction. This comprehensive correction approach demonstrates a high level of efficacy in mitigating the impact of both high and low frequency patterns in the SWIR data. By systematically addressing these spatial artifacts, the overall quality and accuracy of the SWIR imagery are greatly enhanced. This refined data is now well-suited for a wide array of scientific applications, including environmental monitoring, geological studies, and various other forms of Earth observation research.

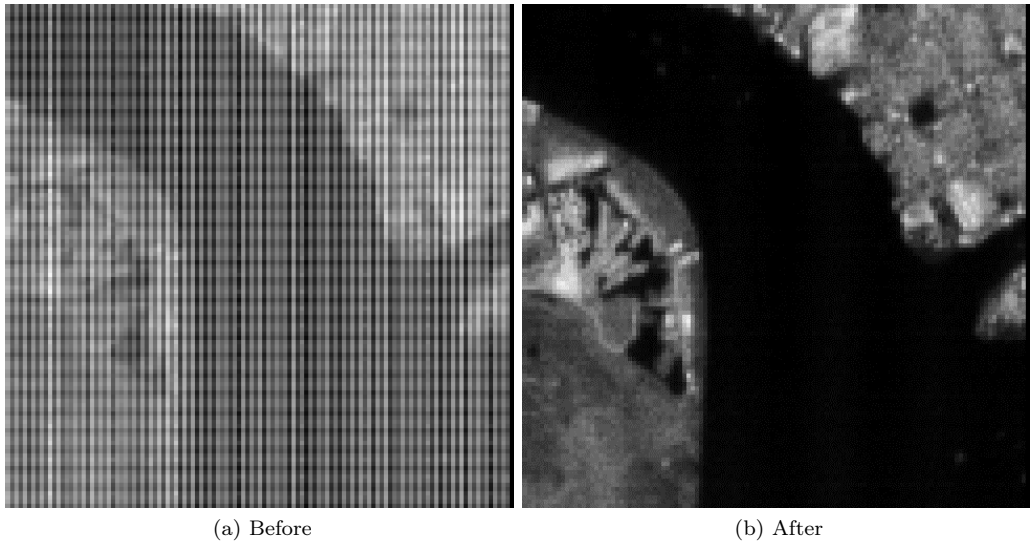

(a) Before  (b) After

Figure 12: Scan Interference & Pattern Noise Correction For Data Quality Improvement

## 4.3 Stray Light Correction

In the SWIR instrument, a significant along-track artifact stemming from out-of-band and out-of-field effects has been identified. This artifact has resulted in both spatial and spectral smearing of the acquired data, severely compromising its overall quality and accuracy. Notably, this along-track smear has been observed across all spectral bands, and the spectral signatures of features are noticeably smoothed. Figure 13 vividly illustrates the impact of this artifact on both moon and earth images. The moon image exhibits clear signs of along-track stray light affecting the overall quality. Additionally, Figure ?? depicts the point spread function derived from the data for band 130, clearly indicating that the along-track smear in the data is more pronounced compared to the across-track direction, extending up to 15 pixels.

To address this issue comprehensively, extensive analysis of moon data was undertaken to quantitatively assess the extent of the artifact. Given that the satellite housing the instruments operates as an agile platform with a step-and-stare mode, a discernible correlation between the artifact and platform steering, as well as spatial location, was observed in images captured over Earth. This insight will be crucial in refining our approach to mitigating this artifact and improving the overall data quality.

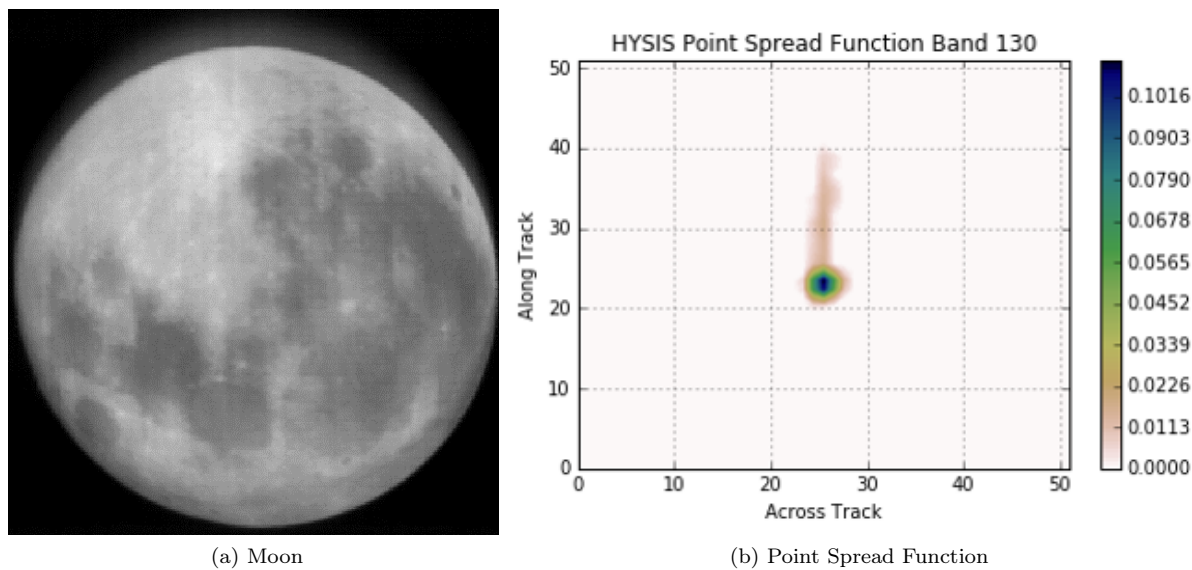

(a) Moon  (b) Point Spread Function

Figure 13: Stray Light Effect in HYSIS

Upon thorough examination of the images acquired over land regions, it has been discerned that the stray light direction is contingent on both platform steering and the spatial location of a pixel. In essence, the direction in which this effect manifests varies based on the specific position of a pixel within the strip. This behavior is vividly demonstrated in Figure 14 for the strip acquired on [date] over [location], where distinct patterns emerge. At the top of the strip, the stray light exhibits a tendency towards the right-hand side. In the center portion, the stray direction appears relatively straight, and as one progresses down the imaging strip, the direction shifts towards the left-hand side. This observation is further corroborated by the estimated point spread functions depicted on the rightmost side of the figures. Understanding this spatial dependency of the stray light with respect to both platform steering and pixel location is a pivotal development in our analysis. It will serve as a cornerstone in formulating targeted strategies to effectively mitigate this artifact and enhance the overall data quality.

Figure 15 underscores that the stray light effect is contingent not only on the position of the feature along the strip but also on the specific pixel location. The initial graph illustrates that for the $100^{th}$ pixel, the stray light direction veers towards the left at the strip start, shifts towards the center at the strip center, and then moves towards the right at the strip end. Similar directional observations are made for the $500^{th}$ and $900^{th}$ pixels, as depicted in the two rightmost graphs. However, it is important to note that the magnitude of the effect varies across these pixels.

Based on these observations, it became evident that the stray light direction exhibited discernible variations at different sections of the strip. Drawing on insights gleaned from platform attitude data, orbital profiles, and a comprehensive understanding of the observed smearing, a robust model was developed. This model was subsequently deployed to rectify the along-track artifact. The correction methodology effectively accounted for the out-of-band and out-of-field effects, resulting in a substantial reduction in smearing and a restoration of the data to a more faithful representation of the true spectral characteristics of the observed targets.

Figures 17 (a), (b), and (c) illustrate image patches corresponding to strip start, center, and end before correction, while (d), (e), and (f) depict the corresponding corrected patches post-application of the algorithm. Figure 16 showcases moon images before and after stray light correction, demonstrating a significant improvement in clarity. Figure ?? demonstrates the effect on city images before and after correction, highlighting a marked increase in sharpness. Figure ?? displays the point spread function before and after correction, with the along-track smear reduced from 15 pixels to 2.5 pixels, indicating a substantial increase in image sharpness. Finally, Figure ?? exhibits the spectral signature of vegetation and sand targets before and after correction, showcasing a noticeable enhancement in spectral contrast and more pronounced dips. This comprehensive correction process has greatly improved the overall

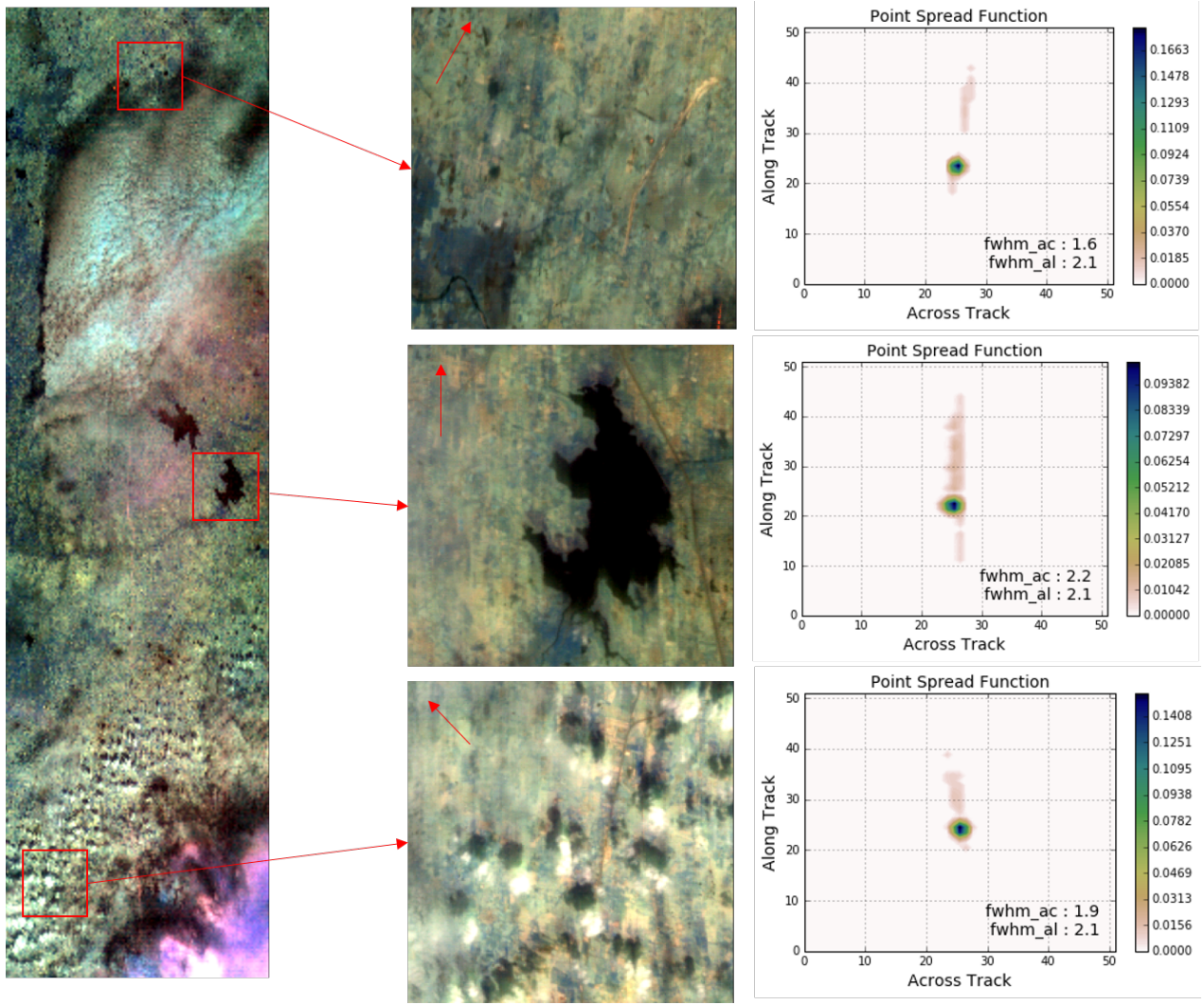

Figure 14: Stray Light Direction being platform steering dependent

quality and accuracy of the data.

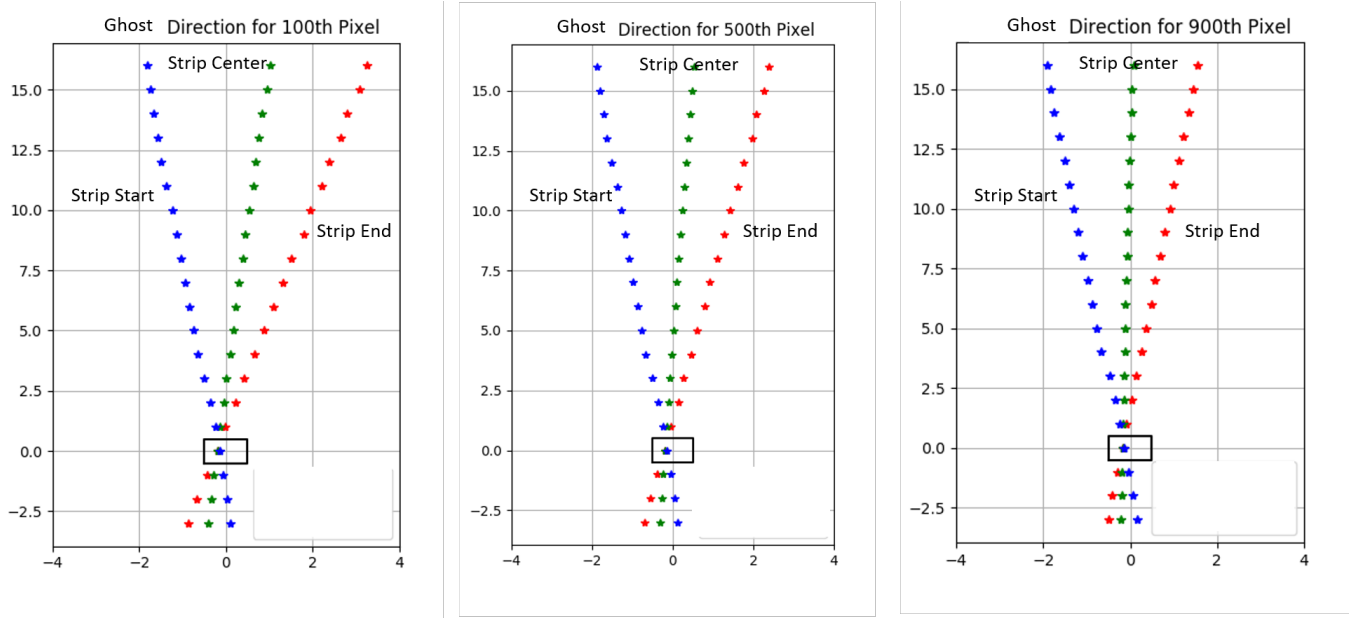

Figure 15: Stray Light Direction being Spatial Pixel dependent

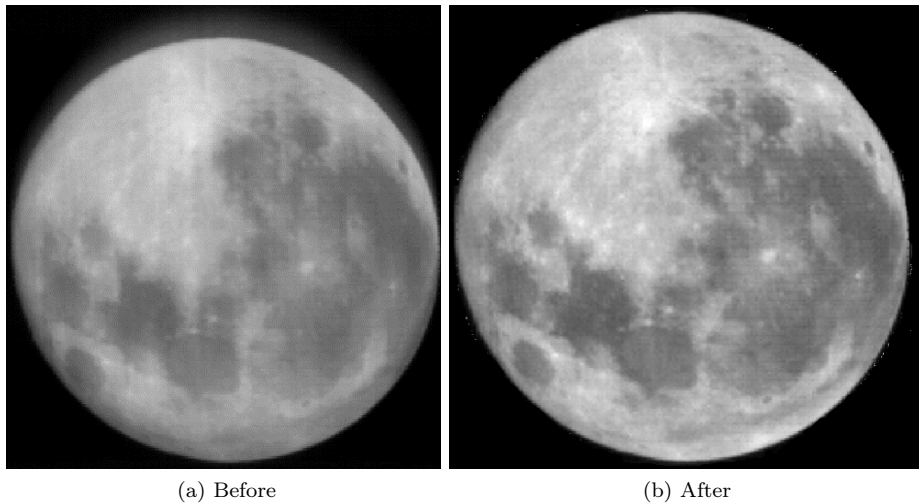

(a) Before  (b) After

Figure 16: Moon Before And After Stray Light Correction

Figure 17 serves as a visual testament to the transformative impact of this correction process. It vividly contrasts the raw data, complete with the artifact, against the corrected images obtained through the newly developed algorithm. The remarkable enhancement in image quality is immediately evident, underscoring the effectiveness of this correction approach. Through the mitigation of the along-track artifact, there's been a substantial boost in the overall quality and accuracy of the data. This correction methodology marks a significant stride forward in guaranteeing the reliability and utility of the SWIR data. By systematically addressing the specific artifact stemming from out-of-band and out-of-field effects, the corrected data now stands as a more faithful representation of the true radiometric values. This enhancement significantly augments its scientific worth across a wide spectrum of applications.

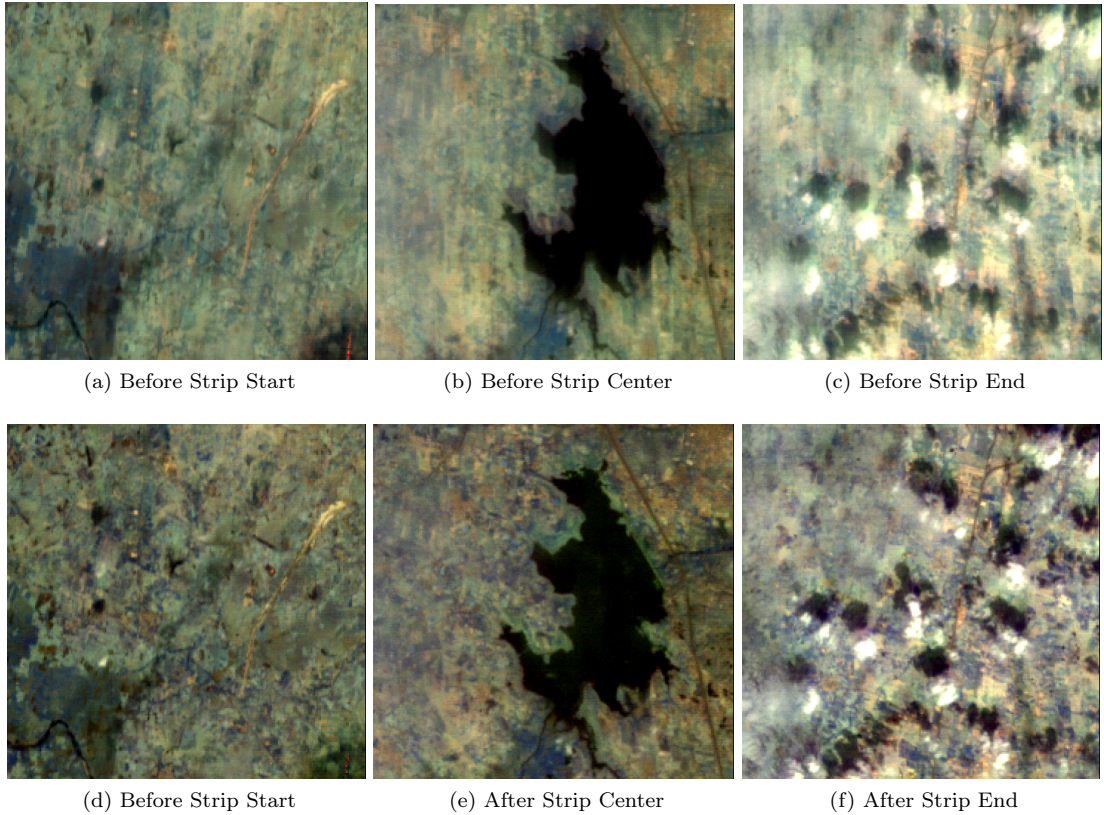

Figure 17: Stray Light Correction

## 5 Inflight Geometric Calibration

Accurate mapping of image plane coordinates (scan and pixel) to precise latitude, longitude, and height on the Earth's surface is a crucial aspect of earth observation satellite missions. This task relies on a combination of detailed information about mounting angles, orbital parameters, and attitude data. The satellite's positioning system, which encompasses both GPS (Global Positioning System) and IRNSS (Indian Regional Navigation Satellite System), plays a vital role in measuring and providing essential orbital parameters. These parameters include information about the satellite's position, velocity, and acceleration, all of which are crucial for accurately determining the satellite's trajectory. Additionally, to ascertain the absolute attitude of the satellite, a combination of data from star sensors and gyroscopes is utilized. Star sensors provide information about the satellite's orientation by detecting the positions of stars in space. Gyroscopes, on the other hand, measure the rate of rotation around different axes, allowing for precise determination of the satellite's attitude. Before the satellite is launched, meticulous measurements are taken to determine the mounting angles of the detectors. These measurements establish the relationship between the detectors and the optical axis, the optical axis and the payload cube, and the payload cube and the master reference cube. These angles are crucial for accurately capturing and recording the data. However, despite these comprehensive preparations, there can be slight deviations from the expected values due to various factors. These may include measurement inaccuracies, post-launch shocks, and thermal effects from both the sun and the release of the satellite from the launch vehicle. To refine and mitigate errors in geo-location, ground control points (GCPs) obtained from reference images are employed. These GCPs serve as known reference points on the Earth's surface with precisely known geographic coordinates. By utilizing these GCPs, any alignment angle biases and focal length variations can be accurately determined and subsequently corrected. This meticulous process results in a substantial improvement in the accuracy of geo-location. Ultimately, this process ensures that the acquired data is precisely and reliably mapped to real-world coordinates on the Earth's surface. This is essential for a wide range of applications, including environmental monitoring, land-use planning, disaster management, and scientific research. The accuracy of geo-location is pivotal in enabling meaningful analysis and interpretation of the satellite imagery.

## 5.1 Residual Mounting Angle Estimation

During the commissioning phase of the mission, a comprehensive assessment of the initial system-level geolocation errors was conducted. For the VNIR instrument, these errors were found to be approximately 3.5 kilometers in the roll direction and 2.0 kilometers in the pitch direction. Conversely, for the SWIR instrument, the errors were approximately 1.5 kilometers in the roll direction and 3.0 kilometers in the pitch direction. To refine and enhance the geolocation accuracy, an in-flight geometric calibration process was initiated. This calibration process involved acquiring a set of 220 strips under varying platform tilt and rate conditions. These strips served as valuable data points for estimating biases in the alignment angles of the instruments relative to the master reference cube. The optimization procedure was executed concurrently for all acquired strips. The primary objective of this optimization was to minimize a defined cost function. This cost function was carefully designed to reduce both the mean error and standard deviation in both across-track and along-track directions, specifically in relation to ground control points derived from Cartosat-1 reference images. Mathematically, this cost function is represented by Equation 1. This calibration process was instrumental in refining the accuracy of the geolocation information, leading to a substantial reduction in the initial system-level errors. By systematically addressing and correcting these geolocation discrepancies, the resulting data provides a more accurate representation of the Earth's surface, facilitating a wide range of scientific applications and analyses. This improved geolocation accuracy is pivotal in ensuring the reliability and utility of the acquired imagery for various earth observation endeavors.

$$Cost = \sum_{}^{strips} (mean_{error}(actual - reference) + std_{error}(actual - reference)) \tag{1}$$

Figure 18 provides a visual representation of the substantial improvements achieved through the calibration exercise. Prior to calibration, the system-level errors were evident, introducing noticeable discrepancies in the geolocation information. However, following the refinement of biases, a remarkable transformation occurred. The system-level error was effectively reduced to a zero mean for both VNIR and SWIR instruments, signifying a significant enhancement in geolocation accuracy. Additionally, the standard deviation, which serves as an indicator of the variability of the errors, witnessed a notable improvement. For the VNIR instrument, the standard deviation decreased to approximately 100 meters in the roll direction and 200 meters in the pitch direction. Similarly, for the SWIR instrument, the standard deviation improved to around 100 meters in the roll direction and 200 meters in the pitch direction. To further elevate the accuracy of the acquired data, an orthorectification process is employed. This process leverages Cartosat-1 reference images and height information derived from a digital elevation model. By aligning and adjusting the data to match the reference images, an impressive level of accuracy is achieved, surpassing that of a single pixel with respect to the reference data. This refined geolocation accuracy is pivotal in ensuring that the acquired imagery is reliable and suitable for a diverse range of applications, including environmental monitoring, urban planning, disaster response, and scientific research. The substantial improvements achieved through the calibration process significantly enhance the data's utility and value for a wide array of earth observation endeavors.

## 5.2 VNIR and SWIR Bundling

The VNIR instrument encompasses a total of 60 channels, enabling the capture of spectral data within the range of 400 nm to 900 nm. On the other hand, the SWIR instrument boasts an impressive 256 bands, allowing for the collection of information spanning from 850 nm to 2500 nm. It's noteworthy that there are 7 bands that are shared between both instruments. To obtain a comprehensive and uninterrupted spectrum covering the extensive range from 400 nm to 2500 nm, it becomes imperative to co-register data from both the VNIR and SWIR instruments. This process involves individual ortho-rectification of VNIR and SWIR data, leveraging Ground Control Points (GCPs) extracted from Cartosat-1 reference images. Additionally, elevation data sourced from a Digital Elevation Model (DEM) is incorporated to enhance the accuracy of the ortho-rectification process. In spite of the meticulous ortho-rectification, there may still be minor residues or discrepancies left uncorrected. To address this, a secondary refinement step is implemented. This step involves establishing correspondences between

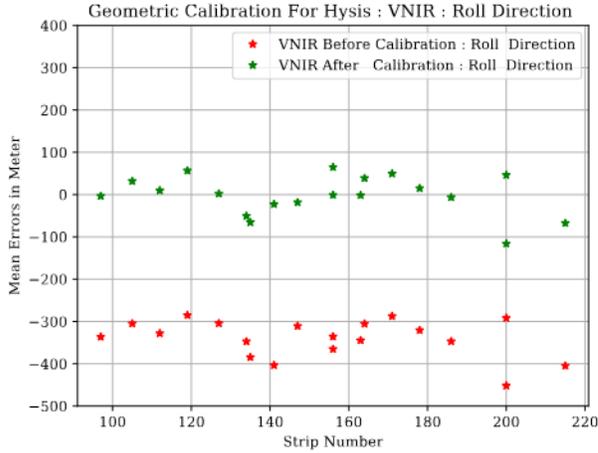
(a) VNIR Roll Direction

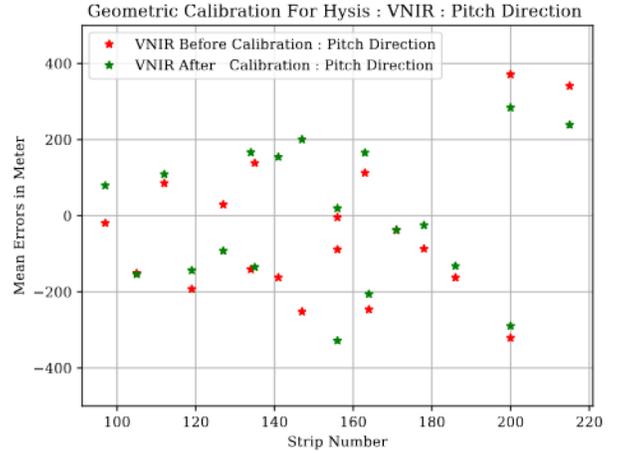
(b) VNIR Pitch Direction

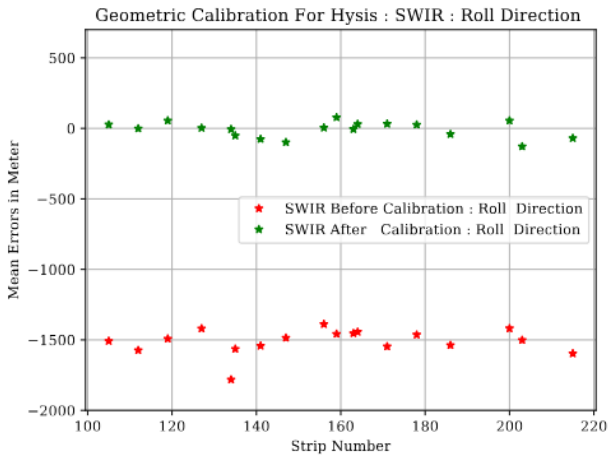
(c) SWIR Roll Direction

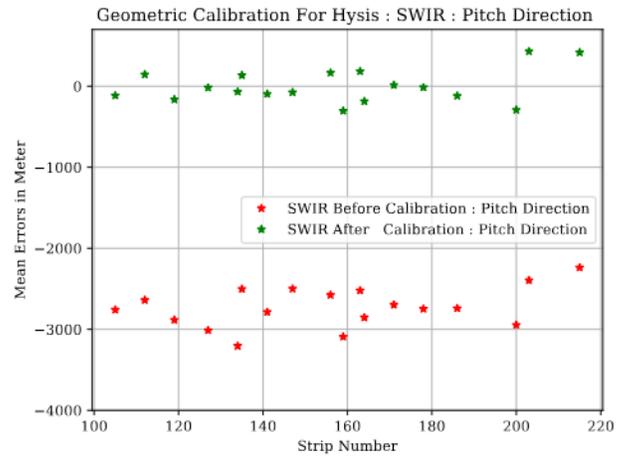
(d) SWIR Pitch Direction

Figure 18: Geometric Calibration Performance Showing Improvement in Geo-location for Both VNIR SWIR Instrument

corresponding bands in the VNIR and SWIR data, effectively identifying any remaining misalignment. Subsequently, an appropriate model is applied to the error surface, enabling precise co-registration of the spectral data from both instruments. The end result is a seamlessly merged spectrum, providing a comprehensive view across the entire range, from 400 nm to 2500 nm, all at an impressive spatial resolution of 30 meters. Figure 19 exemplifies this process over the Okha region in India, presenting both the original data and the continuous spectrum achieved after the bundling process. Importantly, the residual registration errors are consistently maintained at less than 0.25 pixels throughout the entire strip, underscoring the high level of accuracy attained through the bundling process. This comprehensive merging of VNIR and SWIR data significantly enhances the data's utility and value for a wide range of earth observation applications, including environmental monitoring, geological studies, and various other scientific research endeavors. The resulting seamlessly merged spectrum provides a detailed and accurate representation of the Earth's surface across a broad electromagnetic spectrum.

# 6 Conclusion

This paper presents a thorough and systematic account of the calibration procedures meticulously carried out during the pre-launch phase of the Indian hyperspectral mission. The comprehensive

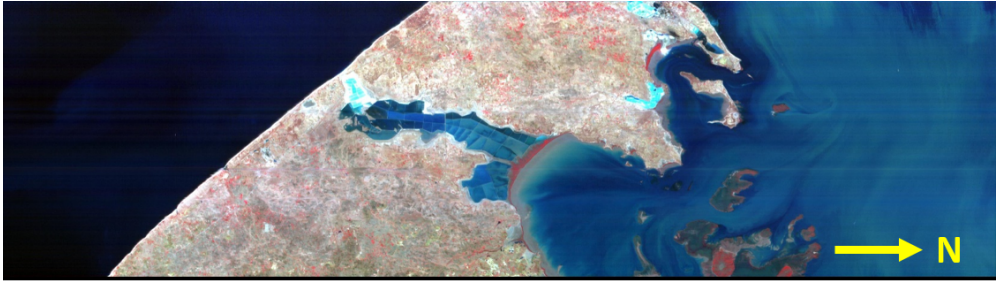

(a) VNIR FCC Combination RGB (B19-B29-B47)

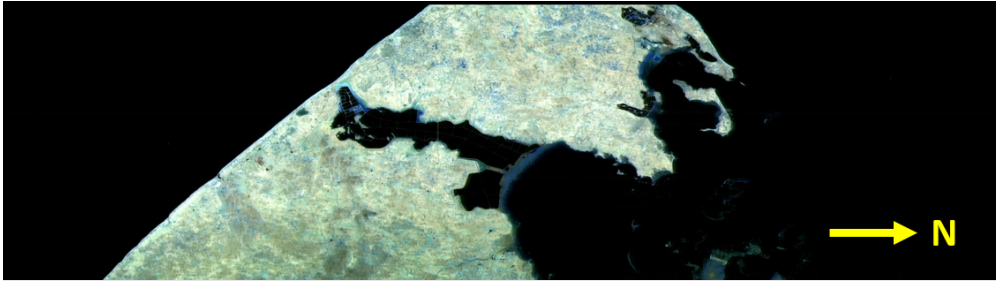

(b) SWIR FCC Combination RGB (B60-B150-B210)

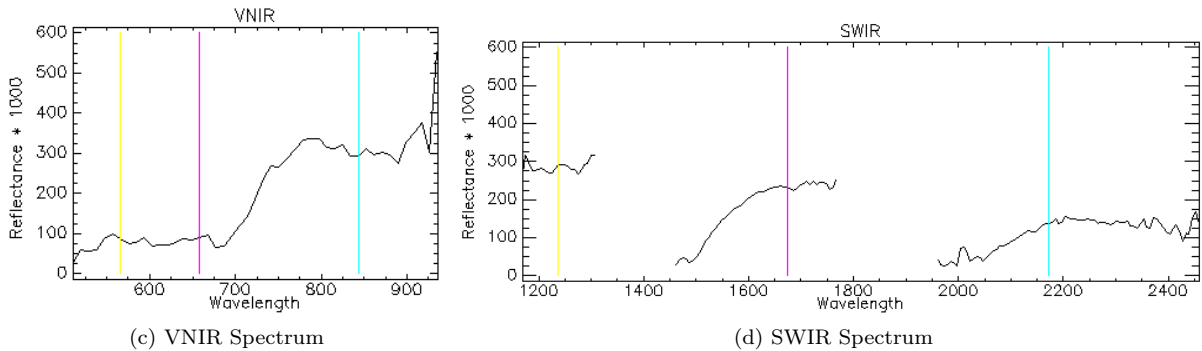

(c) VNIR Spectrum

(d) SWIR Spectrum

Figure 19: Bundled *'Analysis-Ready'* Data Product with sample spectral signature for vegetation, Okha India

evaluation of laboratory performance serves as a crucial foundation, providing essential benchmarks for subsequent in-orbit assessments. The in-depth description of radiometric and geometric calibration efforts in orbit highlights the mission's dedication to achieving optimal data quality. Moreover, the paper not only catalogues the observations made and anomalies encountered in orbit but also goes a step further by elucidating the specialized algorithms developed to rectify these challenges. This innovative approach demonstrates a proactive stance towards ensuring data accuracy and reliability. The delineation of the operational data processing chain, with a focus on the critical stages involved, sheds light on the mission's commitment to generating high-fidelity data products. The assessment of radiometric and geometric accuracy up to the present date provides valuable insights into the operational efficiency and reliability of the satellite in delivering top-notch Earth observations. By implementing refined algorithms, precise calibration coefficients, and robust software modules, the data quality has been elevated to meet stringent standards, culminating in the designation of the resulting data product as *'Analysis-Ready'*. This milestone signifies a pivotal achievement, indicating that the data is now primed for rigorous scientific analysis and application across diverse domains. In summary, this study exemplifies an exemplary approach to calibrating and validating the Indian hyperspectral mission. The seamless integration of pre-launch preparations, comprehensive lab evaluations, and meticulous in-orbit assessments has collectively led to the attainment of a high degree of radiometric and geometric accuracy. The development and application of cutting-edge algorithms underscore the mission's adaptability and problem-solving capabilities. These achievements not only validate the mission's readiness for scientific inquiry but also establish a noteworthy benchmark for future hyperspectral missions worldwide.

# 7 Acknowledgement

Authors would like to thank Director, Space Applications Centre (SAC), Indian Space Research Organisation (ISRO) for the encouragement and guidance provided to carry out this work. Authors also thank members of Signal and Image Processing Area (SIPA) for providing help and support to carry out the activity. Authors also thank Payload Team from SAC, Mission Team from UR Rao Satellite Centre (URSC), Operations and Data Quality and Evaluation (DQE) team of National Remote Sensing Centre (NRSC) and SAC for their help through various stages of the project. Authors also thank SAC reviewers for their valuable suggestions.

# References

[1] Müller, Richard. "Calibration and verification of remote sensing instruments and observations." (2014): 5692-5695.

[2] Thome, K. J. "In-flight intersensor radiometric calibration using vicarious approaches." Post-launch calibration of satellite sensors. CRC Press, 2004. 103-126.

[3] Mouroulis, Pantazis, Robert O. Green, and Thomas G. Chrien. "Design of pushbroom imaging spectrometers for optimum recovery of spectroscopic and spatial information." Applied Optics 39.13 (2000): 2210-2220.

[4] Aktaruzzaman, Md. "Simulation and correction of spectral smile effect and its influence on hyperspectral mapping." ITC, 2008.

[5] Neville, Robert A., Lixin Sun, and Karl Staenz. "Detection of keystone in imaging spectrometer data." Algorithms and Technologies for Multispectral, Hyperspectral, and Ultraspectral Imagery X. Vol. 5425. International Society for Optics and Photonics, 2004.

[6] Foroosh, Hassan, Josiane B. Zerubia, and Marc Berthod. "Extension of phase correlation to subpixel registration." IEEE transactions on image processing 11.3 (2002): 188-200.